\documentclass{aastex}
\usepackage{spr-astr-addons}             
\usepackage{url}
\usepackage{lscape, morefloats, graphicx, float}   
\RequirePackage{color}                   

\begin{document}

\title{CCD $UBV$ photometry of the open cluster NGC~6819}
\slugcomment{Not to appear in Nonlearned J., 45.}
\shorttitle{CCD $UBV$ photometry of the open cluster NGC~6819}
\shortauthors{T. Ak, Z. F. Bostanc\i, T. Yontan, S. Bilir, T. G\"uver, 
S. Ak, H. \"Urg\"up, E. Paunzen}

\author{T. Ak \altaffilmark{1}}
\altaffiltext{1}{Istanbul University, Faculty of Science, Department 
of Astronomy and Space Sciences, 34119 University, Istanbul, Turkey\\
\email{tanselak@istanbul.edu.tr}}

\author{Z. F. Bostanc\i \altaffilmark{1}} 
\altaffiltext{1}{Istanbul University, Faculty of Science, Department 
of Astronomy and Space Sciences, 34119 University, Istanbul, Turkey\\}
\and
\author{T. Yontan \altaffilmark{2}} 
\altaffiltext{1}{Istanbul University, Graduate School of Science and Engineering, 
Department of Astronomy and Space Sciences, 34116, Beyaz\i t, Istanbul, Turkey\\}
\and
\author{S. Bilir \altaffilmark{1}}
\altaffiltext{1}{Istanbul University, Faculty of Science, Department 
of Astronomy and Space Sciences, 34119 University, Istanbul, Turkey\\}
\and
\author{T. G\"uver\altaffilmark{1}} 
\altaffiltext{2}{Istanbul University, Faculty of Science, Department 
of Astronomy and Space Sciences, 34119 University, Istanbul, Turkey\\}
\and
\author{S. Ak\altaffilmark{1}} 
\altaffiltext{2}{Istanbul University, Faculty of Science, Department 
of Astronomy and Space Sciences, 34119 University, Istanbul, Turkey\\}
\and
\author{H. \"Urg\"up \altaffilmark{2}} 
\altaffiltext{1}{Istanbul University, Graduate School of Science and Engineering, 
Department of Astronomy and Space Sciences, 34116, Beyaz\i t, Istanbul, Turkey\\}
\and
\author{E. Paunzen\altaffilmark{3}} 
{\altaffiltext{3}{Department of Theoretical Physics and Astrophysics, 
Masaryk University, Kotl\'a\u rsk\'a 2, 611 37 Brno, Czech Republic\\}

\begin{abstract} We present the results of CCD $UBV$ observations of the open 
cluster NGC~6819. We calculated the stellar density profile in the  cluster's 
field to determine the structural parameters of NGC~6819. Using the existing 
astrometric data, we calculated the probabilities of the stars being physical 
members of the cluster, and used these objects in the determination of the 
astrophysical parameters of NGC~6819. We inferred the reddening and metallicity 
of the cluster as $E(B-V)=0.130\pm0.035$ mag and $[Fe/H]=+0.051\pm 0.020$ dex, 
respectively, using the $U-B$ vs $B-V$ two-colour diagram and UV excesses of 
the F-G type main-sequence stars. We fit the colour-magnitude diagrams of 
NGC~6819 with the PARSEC isochrones and derived the distance modula, distance 
and age of the cluster as $\mu_{V}=12.22\pm 0.10$ mag, $d=2309\pm106$ pc and 
$t=2.4\pm0.2$ Gyr, respectively. The parameters of the galactic orbit estimated 
for NGC~6819 indicate that the cluster is orbiting in a slightly eccentric 
orbit of $e=0.06$ with a period of $P_{orb}= 142$ Myr. The slope of the mass 
function estimated for the cluster is close to the one found for the stars in 
the solar neighbourhood.
\end{abstract}


\section{Introduction}
NGC~6819 ($\alpha_{2000.0}=19^{h}41^{m}18^{s}, \delta_{2000.0}=+40^{\circ}11^{'}12^{''}$; 
WEBDA database\footnote{webda.physics.muni.cz}) is a well-known intermediate 
age ($\sim$2.3 Gyr) open cluster in the Kepler field \citep{Borucki2011}. Its age and 
location in the Kepler field makes it one of the most suitable clusters for investigating 
stellar models and for astroseismic studies of stars in different stages of their 
evolution. The colour excesses, distance moduli, distances, ages, metallicites and 
radial velocities obtained for NGC~6819 in the previous studies are summarized in Table~1.

\begin{table*}[t]
\setlength{\tabcolsep}{5pt}
\begin{center}
\caption{Colour excesses ($E(B-V)$), distance moduli ($\mu_{V}$),
distances ($d$), ages ($t$) and metallicities ($[Fe/H]$) collected from the literature
for the open cluster NGC~6819. References are given in the last column.}
\begin{tabular}{ccccccc}
\hline\hline
$E(B-V)$  & $\mu_{V}$  & $d$  & $t$  & $[Fe/H]$  &  $V_{r}$ & Refs. \\
 (mag)   &  (mag)  & (kpc)  & (Gyr)  & (dex)   & (km~s$^{-1}$) &  \\
\hline\hline
0.12   &  --   & 2.04  & 4.0  & $-0.02\pm0.02$ &  --  & 1  \\
0.30   &  --   & 2.2  &  2  &  --  &  --  & 2  \\
0.28   &  --   & 2.17  &  2  &  --  &  --  & 3  \\
0.16   &  --   &  --  & 2.4  &  $-0.05$  &  --  & 4  \\
$0.14\pm0.04$ &  --   &  --  &  --  & $+0.09\pm0.03$ &  --  & 5  \\
0.10   & $12.30\pm0.12$  & 2.5  &  --  &  --  &  --  & 6  \\
$0.10\pm0.03$ & $12.11\pm0.20$  & 2.3  & $2.5\pm0.6$ &  +0.07  &  --  & 7  \\
0.15   & $12.32\pm0.05$  & 2.3  & $2.5\pm0.6$ &  +0.07  &  --  & 8  \\
$0.14\pm0.04$   & $12.36\pm0.10$  & 2.4  & 2.6  &  +0.09  &  --  & 9  \\
0.15   & $12.20\pm0.06$  & 2.2  & 2.5  &  +0.11  &  --  & 10  \\
0.12   & $12.27\pm0.01$  & 2.4  &  --  &  --  &  --  & 11  \\
$0.160\pm0.007$ & $12.40\pm0.12$  & 2.4  & $2.3\pm0.2$ & $-0.06\pm0.04$ &  --  & 12  \\
--    &  --   &  --  & $2.25\pm0.20$&  --  &  --  & 13  \\
--    &  --   &  --  &  --  &  --  & $-7\pm13$ & 14  \\
--    &  --   &  --  &  --  & $+0.05\pm0.11$ &  --  & 15  \\
--    &  --   &  --  &  --  &  --  & $+4.8\pm0.9$ & 16  \\
--    &  --   &  --  &  --  &  --  & $+1\pm6$ & 17  \\
0.10   &  $12.30$  &  2.5  & 2.4  &  $0$  & $+2.34\pm0.05$ & 18  \\
--    &  --   &  --  &  --  & $-0.02\pm0.02$ & $+2.65\pm1.36$ & 19  \\
\hline\hline
\end{tabular}
\\(1) \cite{Burkhead1971}, (2) \cite{Lind1972}, (3) \cite{Auner1974}, (4) \cite{RV1998}, 
(5) \cite{Bragetal2001}, (6) \cite{Kalietal2001}, (7) \cite{KangAnn2002}, (8) \cite{Basu2011}, 
(9) \cite{Yangetal2013}, (10) \cite{Balo2013}, (11) \cite{Rodetal2014}, 
(12) \cite{AT2014}, (13) \cite{Bedetal2015}, (14) \cite{Frietal1989}, (15) \cite{FriJan1993}, 
(16) \cite{Glusetal1993}, (17) \cite{Thogetal1993}, (18) \cite{THoleetal2009}, (19) \cite{LBetal2015}.
\end{center}
\end{table*}

NGC~6819 has also often been monitored for stellar seismological studies and to search 
for variable stars \citep[e.g.,][]{Stretal2005, Taletal2010, Basu2011, Heketal2011, 
Steletal2011, Migetal2012, Coretal2012, Sandetal2013, Wuetal2014a, Wuetal2014b}. 
Note that \cite{Gosetal2012} discovered X-ray sources within the cluster's half-light 
radius. Although detailed photometric analysis of a cluster is very important to study 
the stellar models, spectroscopic study of the members of a cluster can give valuable 
information for the metallicity and radial velocity of the cluster (see Table~1). 
Recently, \cite{LBetal2015} analyzed high-dispersion spectra of 333 stars in NGC~6819 
to determine the abundances of iron and other metals from spectral features in the 
400 \AA\ region surrounding the Li 6708 \AA\ line. They found its metallicity to be 
$[Fe/H]=-0.02\pm0.02$ dex using a sub-sample restricted to main-sequence and turnoff 
stars.

Here, we report the results of CCD $UBV$ observations of the open cluster NGC~6819 since 
it is located in the Kepler field \citep{Borucki2011} and the mean radial velocity and 
metallicity of the cluster is determined from the high resolution spectra of a 
considerable number of cluster's members brighter than $V\sim 16.7$ mag \citep{LBetal2015}. 
The availability of photometric and spectroscopic data encouraged us to re-calculate the 
cluster's astrophysical and kinematical parameters.

We calculate the membership probabilities of the stars in the cluster's field based on 
their proper motions and the mean radial velocity of the cluster. We find the reddening 
and metallicity of NGC~6819 following two independent methods. We infer its distance 
modulus and age by fitting stellar isochrones to the observed CMDs of the cluster, while 
keeping the reddening and metallicity constant \citep{Yonetal2014,Bosetal2015}. With this 
method, we believe that the parameter degeneracy/indeterminacy in the simultaneous 
statistical solutions \citep[cf.][]{Anders2004, Kingetal2005, Brid2008, deMeule2013, 
Janes2014} of the astrophysical parameters of NGC~6819 can be reduced.

In Section 2, we summarize the observations and reductions. We present the CMDs, structural 
parameters of NGC~6819, and the membership probabilities of the stars in the cluster field 
in Section 3. In section 4, we measure the astrophysical parameters of the cluster. We 
summarize our conclusions in Section 5.

\section{Observations}

CCD $UBV$ observations of the open cluster NGC~6819 were carried out on 18th May 2015 using 
the 1m Ritchey-Chr\'etien telescope (T100) of the T\"UB\.ITAK National Observatory 
(TUG)\footnote{www.tug.tubitak.gov.tr} located in Bak{\i}rl{\i}tepe, Antalya/Turkey. The 
telescope is equipped with an SI~1100 CCD camera (back illuminated, 4k$\times$4k pixels) 
operating at $-90\deg \rm{C}$. Overall the imaging system has a pixel scale of $0.''31$ 
pixel$^{-1}$, resulting in a total field of view of about $21' \times 21'$. The readout 
noise and gain of the CCD are 4.19e$^{-}$ and 0.55~e$^{-}$/ADU, respectively. The field 
of the cluster was observed using both short and long exposures in each filter in order 
to be able to cover the widest possible flux range. Log of observations is given in Table~2. 
The night was moderately photometric with a mean seeing of $1''.6$. A $V$-band image taken 
with an exposure time of 360s is shown in Fig. 1.

\begin{table}
\caption{Log of observations, with exposure times
for each passband. $N$ denotes the number of exposure.} 
\begin{center}
\begin{tabular}{@{}cc@{}}
\hline\hline
Filter & Exp. time (s)$\times$N    \\
 \hline
$U$ & 360$\times$3, 600$\times$3    \\ 
$B$ & 90$\times$5, 360$\times$3    \\
$V$ & 20$\times$5, 90$\times$5, 360$\times$3 \\
\hline\hline
\end{tabular}\label{logobs}
\end{center}
\end{table}

\begin{figure}[h]
\begin{center}
\includegraphics[trim=3cm 1.6cm 4.02cm 0cm, clip=true, scale=0.65]{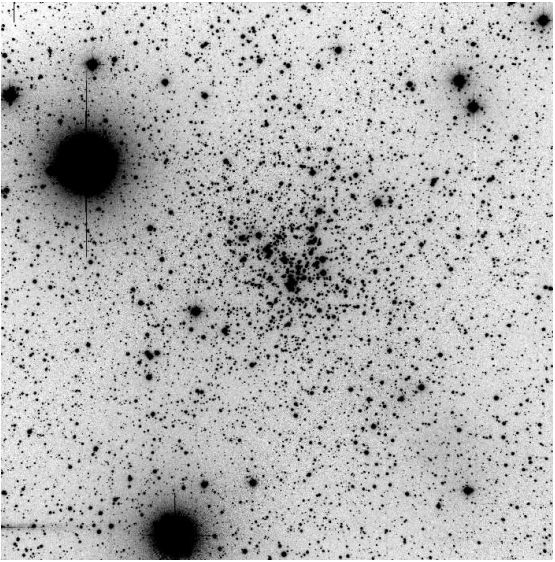}
\caption[] {\small A 360 second $V$-band image of NGC~6819 obtained
with T100 telescope of the T\"UB\.ITAK National Observatory. The image is 
given in inverse colour. The field of view is about 21$\times$21 arcmin 
(North top and East left).}
\end{center}
\end{figure}

For pre-reduction processes and transforming the pixel coordinates to equatorial coordinates, 
Image Reduction and Analysis Facility (IRAF)\footnote{IRAF is distributed by the National 
Optical Astronomy Observatories} routines, PyRAF\footnote{PyRAF is a product of the Space 
Telescope Science Institute, which is operated by AURA for NASA}, and 
astrometry.net\footnote{http://astrometry.net} \citep{Lang2009} software were used together 
with custom written python scripts. Several standard stars selected from \citet{Land2009} 
were also observed during the night to determine the atmospheric extinction and 
transformation coefficients of the observing system. IRAF software packages for aperture 
photometry were used to measure the instrumental magnitudes of the standard stars. Source 
Extractor (SExtractor)\footnote{SExtractor: Software for source extraction} and PSF 
Extractor (PSFEx) \citep{BertArn1996, BertArn2011} together with custom written python 
and IDL scripts were used to detect and measure fluxes of all the objects in the field 
of the cluster and create final source catalogs. Also aperture photometry of a number of 
well separated stars in the field was performed, in order to make aperture corrections to 
the instrumental magnitudes obtained from PSF photometry. Finally, the following equations 
\citep{Janes2011, Janes2013} were used to transform the instrumental magnitudes and colours 
of stars to the standard photometric system:

\begin{equation}
V = v - \alpha_{bv}(B-V)-k_vX _v- C_{bv} \\
\end{equation}

\begin{equation}
B-V = \frac{(b-v)-(k_b-k_v)X_{bv}-(C_b-C_{bv})}{\alpha_b+k'_bX_b-\alpha_{bv}} \\
\end{equation}

\begin{eqnarray}
U-B = \frac{(u-b)-(1-\alpha_b-k'_bX_b)(B-V)}{\alpha_{ub}+k'_uX_u} \\ \nonumber
-\frac{(k_u-k_b)X_{ub}-(C_{ub}-C_b)}{\alpha_{ub}+k'_uX_u} 
\end{eqnarray}
where $U, B$ and $V$ denote the magnitudes in the standard photometric 
system, $u$, $b$ and $v$ the instrumental magnitudes and $X$ the airmass. 
$k$ and $k'$ are primary and secondary extinction coefficients while 
$\alpha$ and $C$ are transformation coefficients to the standard system. 
The photometric extinction and transformation coefficients for that 
particular night were obtained applying multiple linear fits to the 
instrumental magnitudes of the standard stars. The resulting values are 
given in Table 3.

\begin{table*}[t]
\caption{Derived transformation and extinction coefficients. $k$ and $k'$ denote 
primary and secondary extinction coefficients, respectively, while {\it $\alpha$} 
and $C$ are transformation coefficients.} 
\begin{center}
\begin{tabular}{ccccc}
\hline\hline
Band/Coefficient &  $k$  &  $k'$   & {\it $\alpha$} &  $C$  \\
\hline 
   $u$ & $0.429\pm0.053$ & $-0.240\pm0.084$ &  -   &  -  \\
   $b$ & $0.229\pm0.058$ & $0.082\pm0.061$ & $0.741\pm0.097$ & $1.496\pm0.092$ \\
   $v$ & $0.199\pm0.009$ &  -   &  -   &  -  \\
   $u-b$ &  -   &  -   & $1.119\pm0.123$ & $3.852\pm0.081$ \\
   $b-v$ &  -   &  -   & $0.055\pm0.016$ & $1.345\pm0.026$ \\
\hline\hline
\end{tabular}
\end{center}
\end{table*}

\begin{table*}
\setlength{\tabcolsep}{2pt}
\begin{center}
\small{
\caption{Photometric and astrometric catalogue for the open cluster NGC~6819.
The complete table can be obtained electronically. }
\begin{tabular}{ccccccccc}
\hline\hline
 ID & $\alpha_{2000}$ & $\delta_{2000}$ & $V$  &  $U-B$  &  $B-V$  & $\mu_{\alpha}\cos \delta$ & $\mu_{\delta}$ & $P$ \\
   & (hh:mm:ss.ss) &(dd:mm:ss.ss)& (mag) &  (mag) &  (mag)  & (mas yr$^{-1}$)   & (mas yr$^{-1}$) & (\%) \\
\hline
 1 & 19:40:24.59 & 40: 3:18.58 & 20.440$\pm$0.046 &  --  & 1.496$\pm$0.111 & -8.5$\pm$5.5 & -2.2$\pm$5.5 & 85 \\
 2 & 19:40:24.65 & 40: 9:40.85 & 18.709$\pm$0.010 & 0.211$\pm$0.060 & 0.753$\pm$0.018 & -0.6$\pm$7.8 & -0.7$\pm$7.8 & 94 \\
 3 & 19:40:24.66 & 40: 9:58.24 & 19.883$\pm$0.026 &  --  & 1.484$\pm$0.068 & -105.1$\pm$6.2 & -171.7$\pm$6.2 & 0 \\
 4 & 19:40:24.73 & 40:13:41.68 & 20.387$\pm$0.041 &  --  & 1.536$\pm$0.110 &  --  &  --  & -- \\
 5 & 19:40:24.75 & 40: 7: 4.94 & 18.102$\pm$0.006 & 0.540$\pm$0.052 & 0.939$\pm$0.012 & 2.0$\pm$4.0 & 1.5$\pm$4.0 & 44 \\
 -- &  --  &  -- &  --   &  --  &  --  &  --  &  --  & -- \\
 -- &  --  &  -- &  --   &  --  &  --  &  --  &  --  & -- \\
 -- &  --  &  -- &  --   &  --  &  --  &  --  &  --  & -- \\
\hline\hline
\end{tabular}
}
\end{center}
\end{table*}

\section{Data analysis}

\subsection{Identification of stars and photometric errors}
In the field of NGC~6819, we identified 7382 sources and constructed a
photometric and astrometric catalogue. We used the stellarity index
(SI) provided by SExtractor to detect non-stellar objects, most
likely galaxies, in this catalogue. A source with the SI close to 1 is
a point source (most likely a star), while an extended object
has an SI close to zero \citep{BertArn1996}. \cite{Andetal2002} and
\cite{Karetal2004} showed that the objects with an SI smaller than 0.8
can be assumed to be extended objects. We adopted this limit, to 
determine objects which are most likely stars. Resulting catalogue
contains 7060 stars and is given in Table 4. The columns of the table
are organized as ID, equatorial coordinates, apparent magnitude ($V$),
colours ($U-B$, $B-V$), proper motion components ($\mu_{\alpha}\cos
\delta$, $\mu_{\delta}$) and the probability of membership ($P$). The
proper motions of the stars were taken from the astrometric catalogue
of \citet[PPMXL;][]{Roesetal2010}.

The errors of the measurements in the $V$ band and $U-B$ and $B-V$
colours are shown in Fig. 2 as a function of the apparent $V$
magnitude. Mean errors in the selected magnitude ranges are listed in
Table 5. The errors are relatively small for stars with $V<18$ mag,
while they increase exponentially towards fainter magnitudes. As
expected, the largest errors for a given $V$ magnitude occur in the
$U-B$ colours of the stars. For stars brighter than $V=16$ mag and
stars with $V$ magnitudes between 16 and 19, the mean photometric
errors in the $B-V$ colour are smaller than 0.005 and 0.016 
mag, respectively. For the same $V$ band intervals the mean 
errors in the $U-B$ colour index are smaller than 0.009 and 0.065 mag, 
respectively.

\begin{figure}[h]
\begin{center}
\includegraphics[scale=0.82, angle=0]{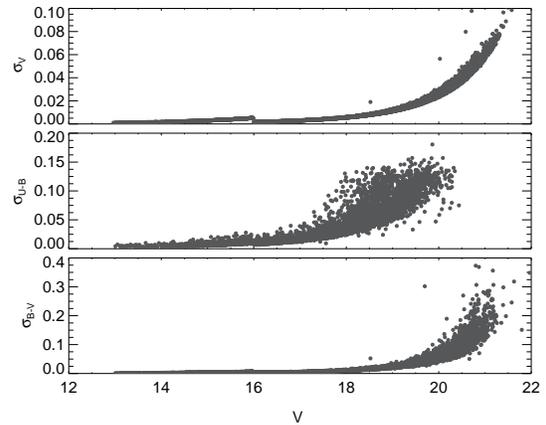}
\caption[] {Colour and magnitude errors of the stars observed in the line-of-sight 
of the open cluster NGC~6819, as a function of $V$ apparent magnitude.} 
\end{center}
\end {figure}

\begin{figure*}[t]
\begin{center}
\includegraphics[scale=0.7, angle=0]{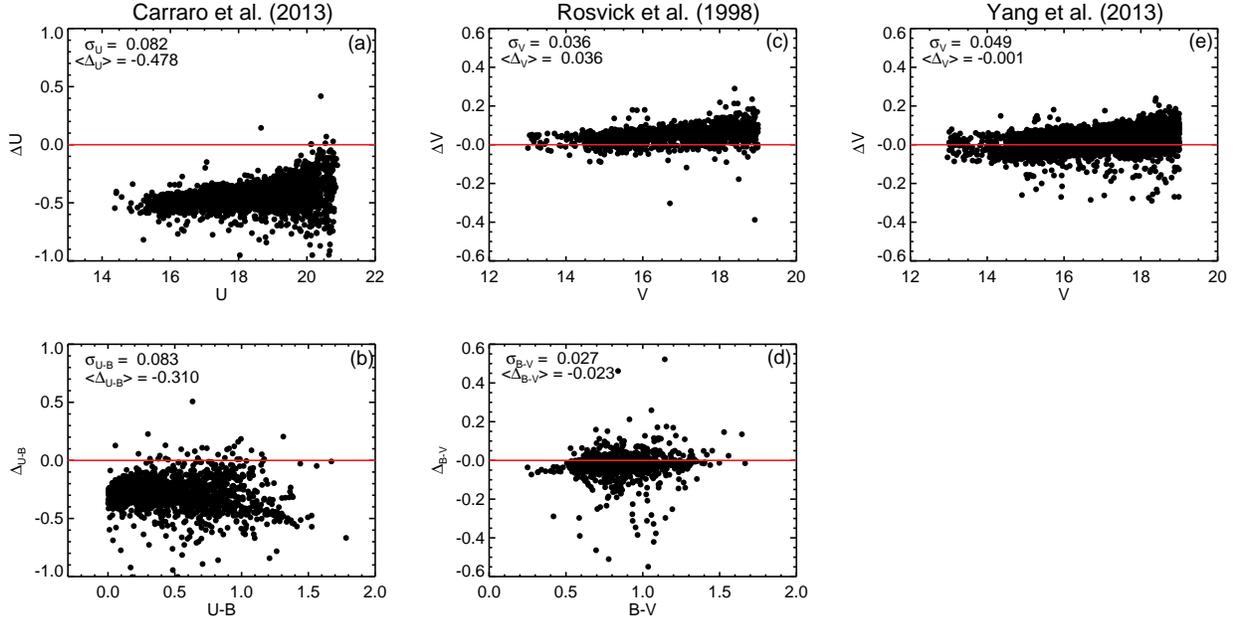}
\caption[] {\small Comparison of the magnitudes and colours in this study 
with those of \cite{Caretal2013} (a and b), \cite{RV1998} (c and d) and 
\cite{Yangetal2013} (e). The means and standard deviations of the differences 
are shown in panels.}
\end{center}
\end{figure*}

\begin{table}[t]
\setlength{\tabcolsep}{4pt}
\caption{Mean errors of the photometric measurements for the stars in the direction
of NGC~6819. $N$ indicates the number of stars within the $V$ apparent magnitude
range given in the first column.}
\begin{center}
\begin{tabular}{ccccc}
\hline\hline
Mag. Range & $N$ & $\sigma_V$ & $\sigma_{U-B}$ & $\sigma_{B-V}$\\
\hline
$12<V\leq14$ & 87 & 0.001 & 0.005 & 0.002 \\
$14<V\leq15$ & 212 & 0.002 & 0.006 & 0.003 \\
$15<V\leq16$ & 572 & 0.004 & 0.009 & 0.005 \\
$16<V\leq17$ & 588 & 0.002 & 0.013 & 0.004 \\
$17<V\leq18$ & 950 & 0.004 & 0.029 & 0.008 \\
$18<V\leq19$ & 1431 & 0.008 & 0.065 & 0.016 \\
$19<V\leq20$ & 1845 & 0.018 & 0.099 & 0.038 \\
$20<V\leq21$ & 1332 & 0.037 & 0.150 & 0.093 \\
$21<V\leq24$ & 43 & 0.120 & --- & 0.212 \\
\hline\hline
\end{tabular}
\end{center}
\end{table}

\begin{table*}[t]
\setlength{\tabcolsep}{4pt}
\caption{Means and standard deviations of the magnitude 
and colour differences between this study and previous studies.}
\begin{center}
\begin{tabular}{cccccccccc}
\hline\hline
\multicolumn{4}{c}{\cite{Caretal2013}}&\multicolumn{4}{c}{\cite{RV1998}}&\multicolumn{2}{c}{\cite{Yangetal2013}} \\
$<\Delta_U>$ & $\sigma_U$ & $<\Delta_{U-B}>$ & $\sigma_{U-B}$ & $<\Delta_V>$ & $\sigma_V$ & $<\Delta_{B-V}>$ & $\sigma_{B-V}$ &
$<\Delta_V>$ & $\sigma_V$ \\
\hline
(mag) & (mag) & (mag) & (mag) & (mag) & (mag) & (mag) & (mag) & (mag) & (mag) \\
\hline
0.082 & -0.478 & 0.083 & -0.310 & 0.036 & 0.036 & 0.027 & -0.023 & 0.049 & -0.001 \\
\hline\hline
\end{tabular}
\end{center}
\end{table*}

We compared our photometric measurements with those of \cite{Caretal2013}, 
\cite{RV1998} and \cite{Yangetal2013} in Fig. 3 using all the stars 
detected both in those observations and ours. In Fig. 3, values on the 
abscissae refer to our measurements, while the magnitude or colour 
differences in the ordinates present the differences between the two 
catalogues. Mean differences and standard deviations obtained for 
each of the five panels are also listed in Table~6. As evident from 
Fig.~3, a good agreement is found from 
the comparison of our measurements with those of \cite{RV1998} and 
\cite{Yangetal2013}. While, the mean magnitude and colour residuals are 
$<\Delta_{V}>=0.036 \pm 0.036$ and $<\Delta_{B-V}>=-0.023\pm 0.027$ mag 
from the comparison with \cite{RV1998}, the mean magnitude residual is 
$<\Delta_{V}>=-0.001\pm 0.049$ mag from the comparison with 
\cite{Yangetal2013}, over the whole star sample. As for the
comparison with \cite{Caretal2013}, differences between the two
catalogues is striking for $\Delta_{U}$ and $\Delta_{U-B}$ with the
mean magnitude and colour residuals of $<\Delta_{U}>=-0.478\pm 0.082$
and $<\Delta_{U-B}>=-0.310\pm 0.083$, respectively, over the whole
star sample. Note that while transforming the instrumental magnitudes 
to the standard system \cite{Caretal2013} did not use a $B-V$ colour 
term, which may be the reason of this discrepancy. In addition, the 
colour excess $E(U-B)=0.15$ estimated in \cite{Caretal2013} is 
considerably higher than expected from the colour excess
($E(B-V)$) values derived in the other studies (see Table 1), since 
the colour excess $E(B-V)$ should be about 0.21 according to the 
relation $E(U-B)/E(B-V)=0.72$ when $E(U-B)=0.15$ mag.

Determination of the photometric completeness limit of the data is
important to calculate reliably the astrophysical and structural
parameters. In order to find the completeness limit of our 
observations in the $V$ band, we constructed a histogram of $V$
magnitudes (see. Fig. 4). From this histogram, we concluded that the
completeness limit of the $V$ magnitudes is 19 mag as this is the mode 
of the distribution. Thus, we decided to only use the stars with 
$V\leq 19$ mag for further analysis. With this selection the number 
of remaining stars for the analysis is 3840 in the field of NGC~6819.

\begin{figure}[h]
\begin{center}
\includegraphics[scale=0.45, angle=0]{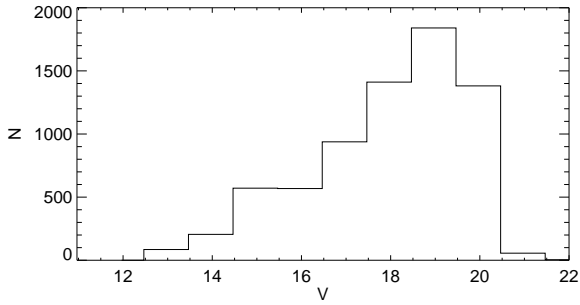}
\caption[] {\small Histogram of the $V$ magnitudes measured in the field of 
the open cluster NGC~6819.}
\end{center}
\end{figure}

\subsection{Cluster radius and radial stellar surface density}

We calculated the structural parameters of NGC~6819 by counting the
number of stars in different annuli around the center of the
cluster. This allowed us to estimate the stellar density profile
of the open cluster NGC~6819 using 3840 stars with $V\leq 19$ mag in
the field. The central coordinates of the cluster were assumed to be
as given by \cite{Yangetal2013} ($\alpha_{2000.0}=19^{h}41^{m}16^{s}.43,
\delta_{2000.0}=+40^{\circ}11^{'}48^{''}.88$). Then, the stellar
density in an area defined by a circle centered on these coordinates 
with a radius of 1.5 arcmin was calculated. From this
central circle, the variation of stellar density using annuli with
widths of 1 arcmin was calculated. The last annulus had a width of
0.75 arcmin because of the decrease in the number of stars. These
calculations of the stellar density were used to plot the stellar
density profile in Fig. 5. We fitted the density profile in Fig. 5
with the \cite{King1962} model defined as,

\begin{equation}
\rho(r)=f_{bg}+\frac{f_0}{1+(r/r_{c})^2},
\end{equation}
where $r$ represents the radius of the cluster centered at the
celestial coordinates given above. $f_{bg}$, $f_0$ and $r_c$ denote
the background stellar density, the central stellar density and the
core radius of the cluster, respectively. In the fitting process, we
used a $\chi^{2}$ minimization technique to determine $f_{bg}$, $f_0$
and $r_c$. The best fit to the density profile is shown with a solid
line in Fig. 5. From this fit, the central stellar density and core
radius of the cluster, together with the background stellar density 
were inferred as $f_{0}=13.18\pm 0.46$ stars arcmin$^{-2}$, 
$r_{c}=3.65\pm 0.38$ arcmin and $f_{bg}=5.98\pm 0.45$ stars 
arcmin$^{-2}$, respectively. The core radius of the cluster derived in 
this study compares with $r_{c}=2.80\pm0.17$ arcmin estimated by 
\cite{Yangetal2013}. Assuming a distance of about 2.3 kpc for the cluster 
(see Table 1), the inferred value corresponds to a core radius of 
$r_{c}=2.44\pm 0.26$ pc, which compares with the core radius of 
$r_{c}=1.75$ pc found by \cite{Kalietal2001}.

\begin{figure}[h]
\begin{center}
\includegraphics[scale=0.42, angle=0]{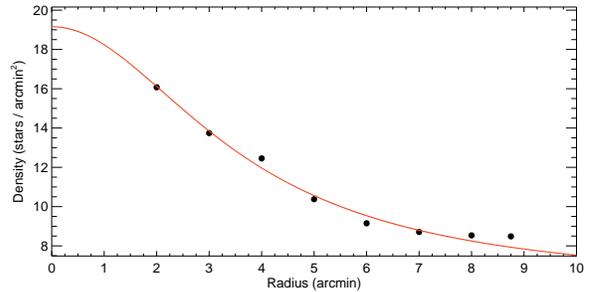}
\caption[] {\small Stellar density profile of NGC~6819.
Errors were determined from sampling statistics:
$1/\sqrt{N}$, where $N$ is the number of stars used in the density
estimation.}
\end{center}
\end{figure}

\subsection{CMDs and membership probabilities}

We constructed $V$ vs $U-B$ and $V$ vs $B-V$ CMDs to derive the
astrophysical parameters of NGC~6819. The CMDs of NGC~6819 are shown 
in Fig. 6. The filled circles in Fig. 6 show the 
248 stars in our field of view, whose high-dispersion spectra 
were analyzed by \cite{LBetal2015}. An inspection by eye suggests
that the cluster is rather dense and its main sequence and giant stars
can be easily distinguished. The turn-off point including a small
group of bright and blue stars in the CMDs of the cluster is located
between $V\sim$14.5 and 15.5 mag. Location of these stars in the CMDs
are very important in the age determination of the cluster.

\begin{figure*}[t]
\begin{center}
\includegraphics[trim=0cm 0cm 0cm 0cm, clip=true, scale=0.7]{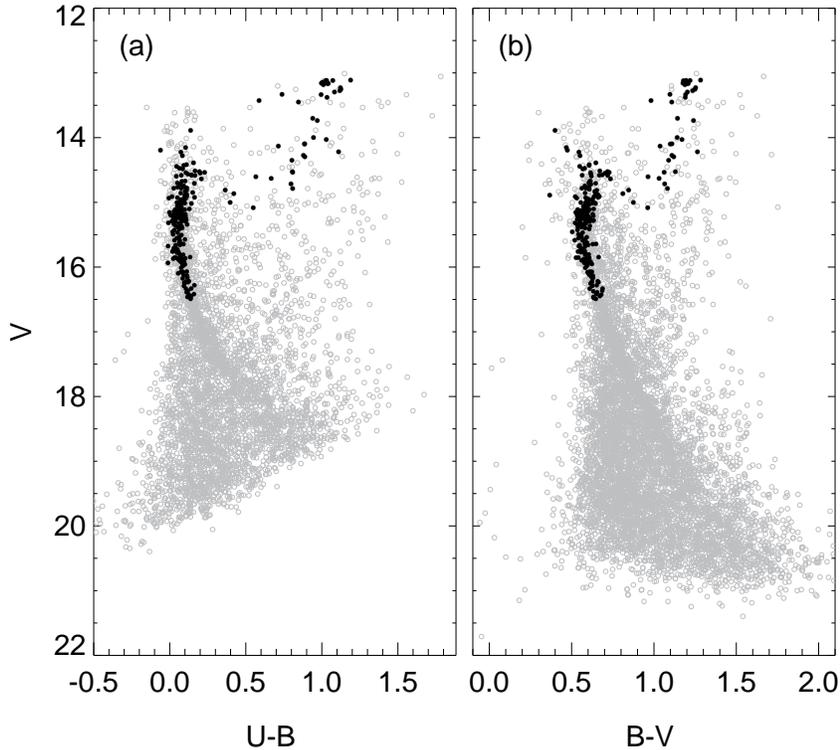}
\caption[] {\small The CMDs for the cluster NGC~6819.
(a) $V$ vs $U-B$ and (b) $V$ vs $B-V$. The filled circles present the stars with 
the high-dispersion spectra analyzed in \cite{LBetal2015}}.
\end{center}
\end{figure*}

Red clump (RC) stars can be used as standard candles for distance 
estimates \citep[i.e.][]{PacSta1998, Cabetal2005, Cabetal2007, 
Biletal2013a, Karaali2013} also their location on a CMD has a crucial 
importance in fitting the theoretical isochrones. The RC stars in 
CMDs of the open clusters can be used to determine the distances 
and ages of them, since the location of the RC stars on the $V$ vs
$B-V$ CMD can be easily found in the colour range $0.7 \leq (B-V)_{0}
\leq 1.2$ mag and the absolute magnitude range $0\leq M_{V} \leq 2$
mag \citep{Biletal2013b}. Here, $(B-V)_{0}$ denotes the de-reddened
$B-V$ colour. Thus, RC stars of the open cluster NGC~6819 should be
located roughly in the same colour range and between the
apparent magnitudes $V\sim 13$ and 13.5 of the $V$ vs $B-V$ CMD of
NGC~6819. An inspection by eye demonstrate that there are stars in
the mentioned colour and apparent magnitude ranges in Fig.~6. Once we
determine membership probabilities of the stars in the direction of
NGC~6819, the RC and turn-off stars can be used to confirm the
distance and age estimation of the cluster.

In order to use the stars near the RC region and turn-off point of the
CMDs, it should be known if they are physical members of the cluster.
Moreover, NGC~6819's main-sequence can be better determined with
the identification of the likely members of the cluster. Thus, we 
calculated the probabilities of the stars in the field being
physical members of the cluster ($P$). In order to do this, we used
the method described by \citet{Bala98}. In this non-parametric
method, both the errors of the mean cluster and the stellar proper
motions are taken into account, and the cluster and field stars'
distributions are empirically determined without any assumption about
their shape. The kernel estimation technique (with a circular Gaussian
kernel function) was used to derive the data distributions. The proper
motions of the stars were taken from \cite{Roesetal2010}. In order to
compare our results with those of the algorithm published by
\citet{Java06}, we considered rectangular coordinates of the stars in
the field, measured in two epochs, first of our observations and
second the ones obtained from \cite{Roesetal2010}. Consequently, we
found excellent agreement. The histogram of the differences
efficiently discriminate the members of the cluster from the
non-members.

\begin{figure}[t]
\begin{center}
\includegraphics[scale=0.8, angle=0]{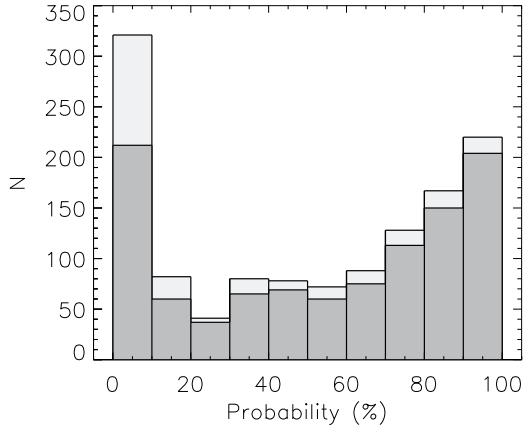}
\caption[] {\small The histogram of the membership 
probabilities of the stars in a circle with a radius of 6 arcmin whose centre 
coincides with NGC~6819's centre for $12 \leq V \leq 19$ (light shaded bars) 
and $15.5 \leq V \leq 19$ mag (shaded bars), respectively.}
\end{center}
\end{figure}

The next step is the determination of the most likely members of
NGC~6819. For this purpose, we first selected the stars in a
circle with a radius of 6~arcmin whose centre coincides with the
cluster's centre. The selected radius of 6 arcmin corresponds to
roughly two times the core radius of NGC~6819, beyond which the field
stars are dominant, as can be seen from Fig.~5. This circle includes
2345 stars within our catalogue. Fig.~7 shows the histogram of the 
membership probabilities of the stars in the selected radius for 
$12\leq V\leq 19$ and $15.5\leq V \leq 19$ mag, respectively. Here, 
as noted above, the magnitude $V=15.5$ roughly corresponds to the 
turn-off point of the cluster, while $V=19$ mag is the photometric 
completeness limit of our measurements. From the histograms in 
Fig.~7, we conclude that the stars with $P\geq 50\%$ are likely 
members of the cluster. We then fitted the zero age main-sequence (ZAMS) 
of \cite{Sungetal2013} for solar metallicity to the $V$ vs $B-V$ CMD 
of NGC~6819 for $15.5\leq V \leq 19$ mag using only the stars with 
$P\geq 50\%$ in order to identify the main-sequence stars of the cluster. 
Note that these stars are also located in the circle 
defined above. By shifting the fitted main-sequence to brighter $V$
magnitudes by 0.75 mag, a band like region in $V$ vs $B-V$ CMD was
obtained to cover the binary stars (see Fig.~8), as well. Hence, we
assumed that all stars with a membership probability $P\geq 50\%$ and
located within the band-like region defined above are the most likely
main-sequence members of NGC~6819, resulting 299 stars. A visual
inspection demonstrate that the stars brighter than $V=15.5$ mag
already left the ZAMS. Thus, we conclude that this magnitude indeed
roughly corresponds to the turn-off point of the cluster. Hence, we
assumed that all stars brighter than $V=15.5$ mag, which are also
located in the circle defined above and have a probability of
membership larger than $P=50\%$, are the most likely members of the
cluster (74 stars). With this procedure, we identified 373 stars for
further analyses which are indicated with red dots in Fig.~8.

\begin{figure}[t]
\begin{center}
\includegraphics[trim=1cm 0cm 0cm 0cm, clip=true, scale=0.55]{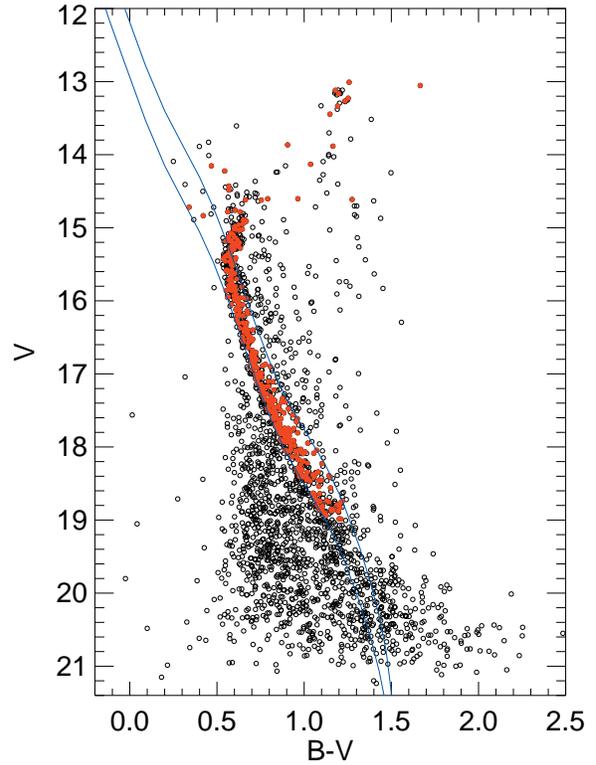}
\caption[] {\small $V$ vs $B-V$ CMD of NGC~6819 constructed
using stars which are located in a circle of 6 arcmin
radius from the centre of the cluster. Solid lines represent the ZAMS
of \cite{Sungetal2013} and the one shifted by an amount of 0.75 mag to
the bright $V$ magnitudes. Red dots indicate the most
probable cluster stars that are identified using a procedure
explained in the text.}
\end{center}
\end{figure}

\section{Determination of the astrophysical parameters of NGC~6819}
\subsection{The reddening}
Reddening of the cluster is the first parameter that must be estimated
since it effects the two-colour diagram (TCD) and CMDs, from which the
remaining astrophysical parameters will be determined. For the
determination of the colour excesses $E(U-B)$ and $E(B-V)$, we used
the most probable 299 main-sequence stars in the 15.5 $\leq V \leq$
19.0 magnitude range, which were selected according to the procedure
in Section 3.3. The positions of these stars in the $U-B$ vs
$B-V$ TCD were compared with the ZAMS of \cite{Sungetal2013} with a 
solar metallicity. In order to do this, based on the proximity 
parameter described in \cite{Jordi96}, we shifted the de-reddened 
main-sequence curve of \cite{Sungetal2013} with steps of 0.001~mag 
within the range $0\leq E(B-V)\leq 0.20$ mag until the best fit is 
obtained with the $U-B$ vs $B-V$ TCD of NGC~6819. We calculated the 
shift in the $U-B$ axis by adopting the following equation 
\citep{HJ53,Gar88}:

\begin{equation}
E(U-B) = E(B-V) \times [0.72 + 0.05 \times E(B-V)].
\end{equation}
We show the $U-B$ vs $B-V$ TCD of NGC~6819 for the most probable 
main-sequence stars of the cluster in Fig.~9. Using this method, we 
estimate the following colour excesses: $E(U-B)=0.094 \pm 0.025$ 
and $E(B-V)=0.130 \pm 0.035$ mag. The errors were estimated by 
shifting the best fit curve for $\pm1\sigma$.

\begin{figure}[t]
\begin{center}
\includegraphics[scale=0.7, angle=0]{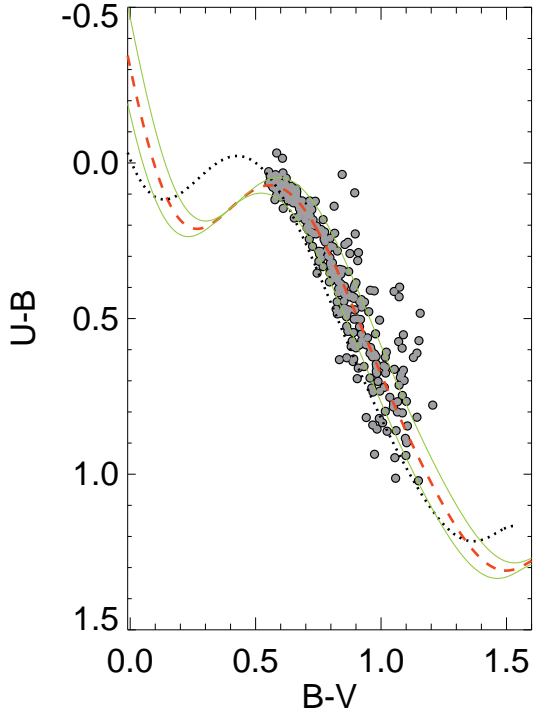}
\caption[] {\small $U-B$ vs $B-V$ TCD for the main-sequence 
stars with $15.5\leq V\leq 19$ mag of NGC~6819. The reddened and 
de-reddened main-sequence curves \citep{Sungetal2013} fitted to the 
cluster stars are represented with red dashed and black dotted lines, 
respectively. Green lines represent $\pm 1\sigma$ deviations.}
\end{center}
\end{figure}

\subsection{Photometric metallicity of NGC~6819}

To measure the photometric metallicity of the open cluster NGC~6819,
we used the method described in \cite{Karetal2011}. Since this
procedure uses F-G type main-sequence stars, we selected 141 of
299 stars with colours $0.3\leq (B-V)_0\leq 0.6$ mag corresponding to
F0-G0 spectral type main-sequence stars \citep{Cox2000}.

The normalized ultraviolet (UV) excesses of the selected stars must be
calculated to utilize the method described in \cite{Karetal2011}. The
normalized UV excess of a star is defined as the difference between
its de-reddened $(U-B)_0$ colour indice and the one corresponding to
the members of the Hyades cluster with the same de-reddened $(B-V)_0$
colour index, i.e. $\delta=(U-B)_{0,H}-(U-B)_{0,S}$. Here, the
subscripts $H$ and $S$ refer to Hyades and star, respectively. 
 Therefore, we calculated the normalized UV excesses of the 141
stars selected as described above and normalized their $\delta$
differences to the UV-excess at $(B-V)_{0}=0.6$ mag,
i.e. $\delta_{0.6}$.

\begin{figure}[t]
\begin{center}
\includegraphics[scale=0.60, angle=0]{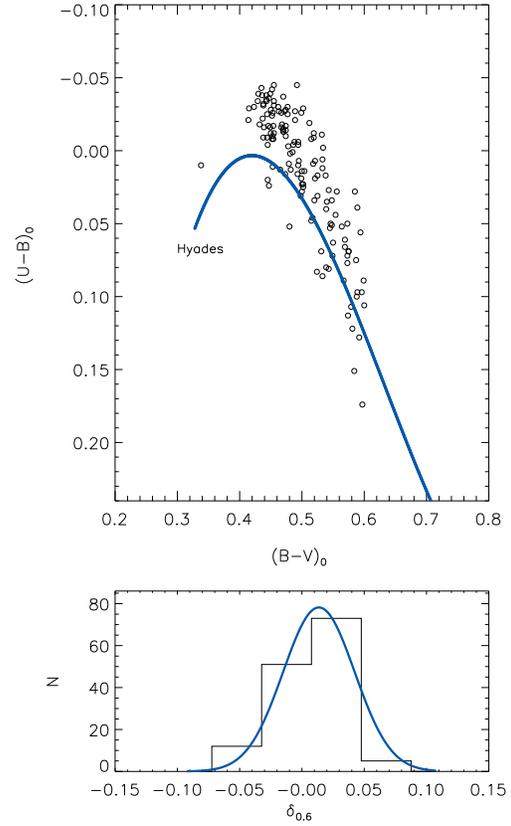}
\caption[] {\small The $(U-B)_{0}$ vs $(B-V)_{0}$ TCD ({\it upper panel}) 
and the histogram ({\it lower panel}) for the normalized UV-excess for 141 
main-sequence stars used for the metallicity estimation of NGC~6819. The solid 
lines in the upper and lower panels represent the main-sequence of Hyades cluster 
and the Gaussian fit of the histogram, respectively.}
\end{center}
\end{figure}

Fig.~10 shows the $(U-B)_0$ vs $(B-V)_0$ TCD and the histogram of the
normalized $\delta_{0.6}$ UV excesses of the selected 141
main-sequence stars of NGC~6819. We calculated the normalized UV
excess as $\delta_{0.6}=0.014\pm 0.002$ mag by fitting a Gaussian to
this histogram, which is also shown in Fig.~10. Here, the
uncertainty is given as the statistical uncertainty of the peak of the
Gaussian. Then, we estimated the metallicity ($[Fe/H]$) of the cluster
by evaluating this Gaussian peak value in the following equation of
\cite{Karetal2011}:

\begin{eqnarray}
[Fe/H]=-14.316(1.919)\delta_{0.6}^2-3.557(0.285)\delta_{0.6}\\ \nonumber
+0.105(0.039).
\end{eqnarray}

The metallicity corresponding to the peak value for the $\delta_{0.6}$
distribution was calculated as $[Fe/H]= +0.051\pm 0.020$ dex. The error 
value of the metallicity was estimated due to the stated errors in the 
colour excess $E(B-V)$ by assuming a colour excess that is 1$\sigma$ 
higher and 1$\sigma$ lower and then calculating $[Fe/H]$ in each case.

The following relation to transform the $[Fe/H]$ metallicities
obtained from the photometry to the mass fraction $Z$ \citep{Mowlavietal2012}:

\begin{equation}
Z=\frac{0.013}{0.04+10^{-[Fe/H]}}.
\end{equation}
Here, $Z$ is the mass fraction of all elements heavier than helium,
which is used to estimate the theoretical stellar evolutionary
isochrones. Hence, we calculated $Z=0.011\pm0.002$ from the metallicity
($[Fe/H]=+0.051\pm0.020$ dex) obtained from the photometry. Since this
abundance is very close to the solar value, which is given as
$Z=0.0152$ by \cite{Bresetal2012}, we prefer to use the solar
abundance in the determination of the astrophysical parameters 
of the cluster.

In order to compare the photometric and spectroscopic metallicities,
we selected the F0-G0 type main-sequence stars in the sample given by
\cite{LBetal2015}, who analyzed high-dispersion spectra of 333 stars
in the field of NGC~6819 to determine the abundances of iron and other
metals and found the cluster's metallicity to be $[Fe/H]=-0.02\pm0.02$
dex using a sub-sample restricted to main-sequence and turnoff
stars. We identified 141 F0-G0 type main-sequence stars in their
sample, for which the mod value of the metallicities is
$[Fe/H]=+0.025$ dex, which is in agreement with the metallicity of
$[Fe/H]=+0.051\pm 0.020$ dex in this study.

\begin{table*}
\setlength{\tabcolsep}{3pt}
\caption{Colour excesses, metallicities ($Z$), distance moduli ($\mu$), 
distances ($d$) and ages ($t$) estimated using two CMDs.}
\begin{center}
\begin{tabular}{lccccc}
\hline\hline
CMD   & Colour Excess & $Z$ & $\mu_{V}$ & $d$ & $t$ \\
    &  (mag) &  & (mag) & (pc) & (Gyr) \\
\hline
$V$ vs $U-B$ & $E(U-B)=0.094\pm0.025$ & $0.0152$ & $12.22\pm 0.10$ & $2309\pm 106$ & $2.4\pm0.2$ \\
$V$ vs $B-V$ & $E(B-V)=0.130\pm0.035$ & $0.0152$ & $12.22\pm 0.10$ & $2309\pm 106$ & $2.4\pm0.2$ \\
\hline\hline
\end{tabular}
\end{center}
\end{table*}

\begin{table*}
\setlength{\tabcolsep}{4pt}
\caption{The RC stars with a membership probability $P\geq 50\%$ in the 
field of NGC~6819. ID number, equatorial coordinates, $V$ magnitude, $B-V$ colour 
and the membership probability of the stars were taken from the main photometric 
catalogue in this study. The $J$-band magnitudes were taken from the 2MASS All-Sky 
Catalog of Point Sources \citep{Cutri03}. Absolute magnitudes were estimated using 
the calibration given by \citep{Biletal2013a}.}
\begin{center}
\begin{tabular}{ccccccccc}
\hline\hline
ID & $\alpha_{2000}$ & $\delta_{2000}$ &  $V$   & $J$   & $B-V$   & $P$ & $M_V$ & $d$  \\
  &(hh:mm:ss.ss) & (dd:mm:ss.ss) & (mag)   &  (mag)  & (mag)   & (\%)  & (mag) & (pc)  \\
\hline
1417 & 19:40:50.24 & +40:13:10.92 & 13.120$\pm$0.001 & 11.009$\pm$0.020 & 1.179$\pm$0.002 & 51 & 0.922 & 2286$\pm$39 \\
1858 & 19:40:57.06 & +40:10:06.93 & 13.335$\pm$0.001 & 11.155$\pm$0.020 & 1.192$\pm$0.002 & 85 & 0.930 & 2514$\pm$42 \\
1930 & 19:40:57.89 & +40:13:53.17 & 13.009$\pm$0.001 & 10.775$\pm$0.020 & 1.258$\pm$0.002 & 95 & 0.972 & 2123$\pm$36 \\
2761 & 19:41:09.30 & +40:14:43.64 & 13.164$\pm$0.001 & 11.005$\pm$0.020 & 1.198$\pm$0.002 & 52 & 0.934 & 2320$\pm$40 \\
3024 & 19:41:13.17 & +40:06:42.04 & 13.446$\pm$0.001 & 11.374$\pm$0.020 & 1.148$\pm$0.002 & 93 & 0.903 & 2680$\pm$45 \\
4586 & 19:41:33.31 & +40:12:35.05 & 13.228$\pm$0.001 & 10.935$\pm$0.021 & 1.253$\pm$0.002 & 74 & 0.968 & 2351$\pm$40 \\
4837 & 19:41:36.82 & +40:09:03.31 & 13.266$\pm$0.001 & 10.998$\pm$0.021 & 1.234$\pm$0.002 & 98 & 0.957 & 2406$\pm$40 \\
\hline\hline
\end{tabular}
\end{center}
\end{table*}

\begin{figure*}
\begin{center}
\includegraphics[scale=0.8, angle=0]{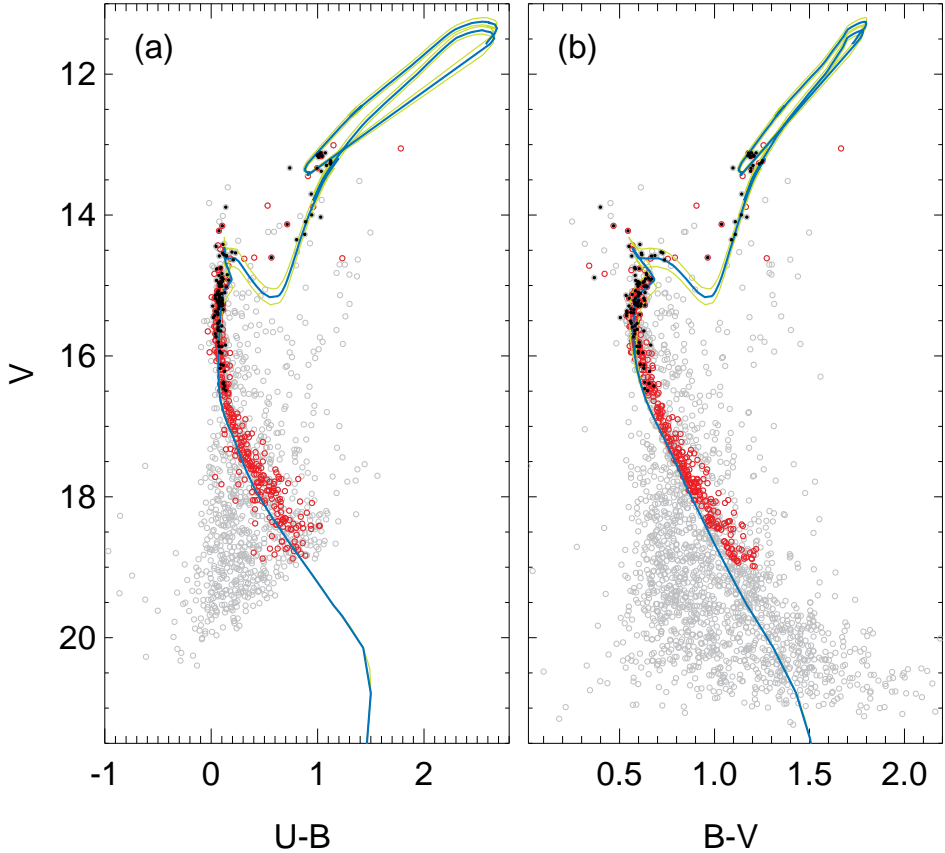}
\caption[] {\small $V$ vs $B-V$ and $V$ vs $U-B$ CMDs for the stars 
located in a circle of 6 arcmin radius from the centre of NGC~6819. 
The most probable members of the cluster are indicated with red circles. 
These stars are fitted to the isochrone determined in this study (blue line). 
The green lines indicate the isochrones with estimated age plus/minus its error. 
Stars indicated with black dotes are the ones with high-dispersion spectra 
analyzed in \cite{LBetal2015}.}
\end{center}
\end{figure*}

\subsection{Distance modulus and age of NGC~6819}

We already measured the reddening and metallicity of the cluster using
its $(U-B)_0$ vs $(B-V)_0$ TCD and the normalized ultraviolet (UV)
excesses of the cluster members, respectively. In order to derive the
distance modula and age of NGC~6819 simultaneously, we fitted the CMDs
of the cluster with the theoretical isochrones provided by the PARSEC
synthetic stellar library \citep{Bresetal2012}, which was recently
updated \citep[PARSEC version 1.2S,][]{Tangetal2014,Chenetal2014}.
The metallicity and the reddening of the cluster were kept
 constant during the fitting process. As already noted, large
uncertainties in the measured reddening, metallicity, and therefore
the age values are marked due to the degeneracies between the
parameters when these parameters are determined simultaneously. Thus,
we ensured that the degeneracy/indeterminacy of the parameters will be
less than that in the statistical solutions with four free
astrophysical parameters (i.e. the metallicity, reddening, distance
modulus and age) by keeping the metallicity and reddening of the
cluster as constants. In Fig.~11, we overplot the best fit
theoretical isochrones for $Z=0.0154$ and $t=2.4~\rm{Gyr}$ in the $V$
vs $B-V$ and $V$ vs $U-B$ CMDs. The estimated astrophysical
parameters of NGC~6819 obtained from the best fits to the CMDs are
given in Table~7. Errors of the parameters were derived by visually
shifting the theoretical isochrones to include all the main-sequence
stars in the CMDs.

\subsection{Distance via the Red Clump Stars}

In order to confirm the distance estimate of the cluster given above,
we utilized the red clump (RC) stars identified in the CMDs since the
RC stars can be used as standard candles in distance estimation of the
objects associated with them \citep[cf.][]{Alves00, Grocholski02, 
Groenewegen08,Yaz13,Biletal2013a,Biletal2013b}. Because the absolute 
magnitude of the RC stars are metallicity dependent especially 
in the optical bands, \citet{Biletal2013a} developed a $V$-band 
absolute magnitude calibration for the RC stars based on their $B-V$ 
colour and metallicity using the RC stars detected in a number of 
stellar clusters, as following,

\begin{eqnarray}
M_{V}=0.627(\pm 0.104)(B-V)_{0}+0.046(\pm 0.043)[Fe/H]\\ \nonumber
+0.262(\pm0.111),
\end{eqnarray}
The calibration equation is valid in the ranges $0.42<(B-V)_{0}<1.20$ mag,
$-1.55<[Fe/H]<+0.40$ dex and $0.43<M_{V}<1.03$ mag.

The RC stars in the $V$ vs $B-V$ CMD of NGC~6819 occupy a region which
is limited with $1.05\leq B-V\leq 1.35$ and $13.0\leq V\leq
13.5$. This region is indicated with a box in Fig. 12a and includes 19
stars. Seven of these stars with $P\geq 50\%$ are denoted with
red circles in Fig.~12. We constructed the $V_0$ vs $J_0$
two-magnitude diagram of these RC candidate stars in order to 
find whether they lie in the giant region defined in
\citet{Bilir06}. Here $J$ denotes the $J$-band apparent magnitude in
the Two Micron All Sky Survey \cite[2MASS;][]{Skrutskie06} photometric
system. The equation set given by \cite{Bilir08} was used to estimate
the de-reddened $J_{0}$ magnitude. Indeed, positions of the selected
stars in the associated diagram (Fig.~12b) suggest that they are
evolved stars \citep[see also,][]{Bilir06b, Bilir10, Bilir11}. We 
estimated their distances adopting the absolute magnitude calibration 
given above \citep{Biletal2013a}, and using the metallicity 
$[Fe/H]=+0.051 \pm 0.020$ dex and colour excess $E(B-V)=0.130 \pm 0.035$ 
mag found in this study. Resulting distances for the RC stars are given 
in Table~8. The mean distance of the RC stars in Table~8 is $d=2383\pm
177$~pc in agreement with the distance obtained from the theoretical
isochrone fits in Table~7, $d=2309\pm 106$~pc. The error value given
for the mean distance obtained from the RC stars is the standard
deviation of the individual distance estimates of the stars. Thus, we
confirm the distance of the cluster estimated by the procedure in
Section 4.3 via seven RC stars of the cluster.

\begin{figure}
\begin{center}
\includegraphics[trim=3cm 0cm 3cm 1cm, clip=true, scale=0.55]{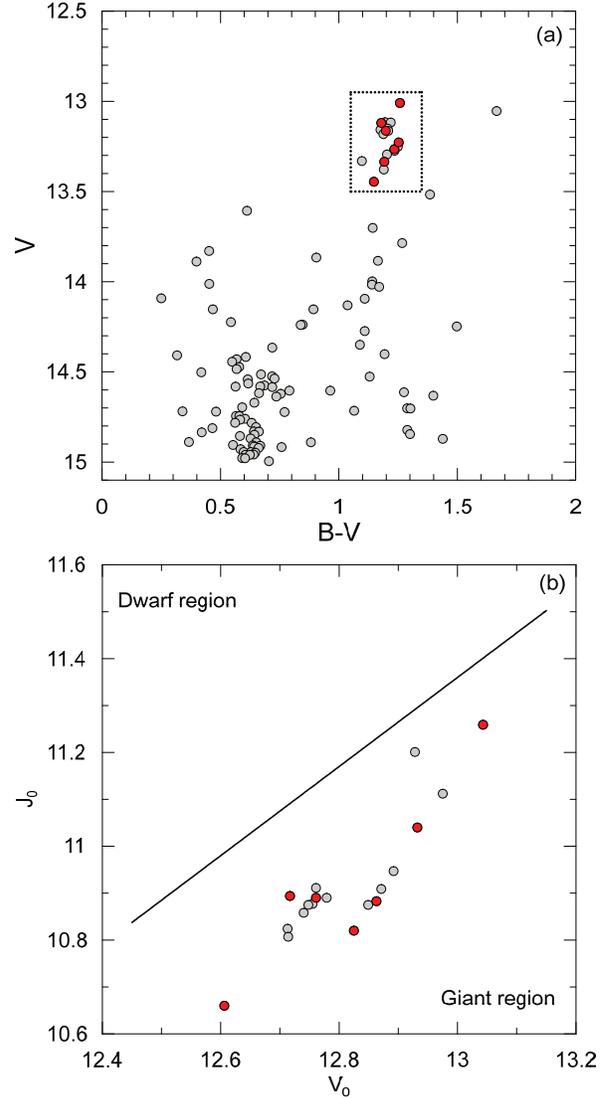}
\caption[] {Location of the RC stars in the $V$ vs $B-V$ CMD of NGC~6819 
indicated with a box (a) and their positions in the $J_0$ vs $V_0$ two-magnitude 
diagram (b). The solid line, $J_0=0.957\times V_0-1.079$, separates the stars 
into dwarf and giant categories \citep{Bilir06}. The red circles indicate 
the seven RC stars with $P\geq 50\%$ in both panels.}
\end{center}  
\end {figure}

\begin{table*}[t]
\setlength{\tabcolsep}{3pt}
\caption{The data for stars used in the calculation of the galactic orbit of NGC~6819.}
\begin{center}
\begin{tabular}{cccccccccc}
\hline\hline
WOCS & ID & $\alpha_{2000}$ & $\delta_{2000}$ & $V$ & $B-V$ & $\mu_{\alpha}\cos\delta$ & $\mu_{\delta}$ & $V_{r}$ & $P$ \\
  &  & (hh:mm:ss.ss) & (dd:mm:ss.ss) & (mag)  & (mag) & (mas~yr$^{-1}$) & (mas~yr$^{-1}$) & (km~s$^{-1}$) & ($\%$)\\
\hline
28012 & 1806 & 19:40:56.46 & +40:07:40.96 & 15.412$\pm$0.004 & 0.607$\pm$0.005 & -4.1$\pm$3.9 & -2.9$\pm$3.9 & 2.3$\pm$1.1 & 100 \\
18013 & 1885 & 19:40:57.37 & +40:16:27.48 & 14.760$\pm$0.003 & 0.604$\pm$0.003 & -4.4$\pm$3.9 & -1.8$\pm$3.9 & 3.4$\pm$1.2 & 97 \\
60012 & 2275 & 19:41:02.67 & +40:16:31.95 & 16.462$\pm$0.002 & 0.632$\pm$0.004 & -3.2$\pm$4.1 & -2.2$\pm$4.1 & 2.0$\pm$1.5 & 98 \\
43009 & 2486 & 19:41:05.51 & +40:08:28.59 & 15.858$\pm$0.005 & 0.607$\pm$0.006 & -4.9$\pm$3.9 & -0.7$\pm$3.9 & 0.7$\pm$1.6 & 90 \\
25005 & 2517 & 19:41:05.87 & +40:12:30.34 & 15.295$\pm$0.004 & 0.598$\pm$0.004 & -3.4$\pm$3.8 & -2.4$\pm$3.8 & 3.1$\pm$1.1 & 98 \\
35008 & 3078 & 19:41:13.94 & +40:15:30.31 & 15.639$\pm$0.004 & 0.580$\pm$0.005 & -2.0$\pm$3.9 & -4.7$\pm$3.9 & 3.8$\pm$1.7 & 92 \\
22007 & 3369 & 19:41:17.86 & +40:15:11.56 & 15.019$\pm$0.003 & 0.645$\pm$0.004 & -3.4$\pm$3.8 & -4.9$\pm$3.8 & 2.8$\pm$1.1 & 96 \\
24007 & 4028 & 19:41:26.18 & +40:14:36.72 & 15.130$\pm$0.003 & 0.600$\pm$0.004 & -3.8$\pm$6.6 & 0.9$\pm$6.6 & 2.9$\pm$3.4 & 91 \\
25008 & 4774 & 19:41:35.86 & +40:10:27.72 & 15.388$\pm$0.004 & 0.573$\pm$0.005 & -5.5$\pm$5.1 & -1.9$\pm$5.1 & -0.2$\pm$1.1 & 96 \\
39009 & 4913 & 19:41:38.02 & +40:10:50.26 & 15.805$\pm$0.005 & 0.604$\pm$0.006 & -3.7$\pm$3.9 & -0.7$\pm$3.9 & 2.9$\pm$1.0 & 90 \\
31011 & 5435 & 19:41:45.27 & +40:10:54.17 & 15.687$\pm$0.004 & 0.580$\pm$0.005 & -3.8$\pm$3.9 & -2.1$\pm$3.9 & 3.9$\pm$1.5 & 98 \\
\hline\hline
\end{tabular}
\end{center}
\end{table*}

\subsection{Galactic orbit of the cluster}

We estimated the parameters of the galactic orbit of the cluster
following  the  procedure  described in \cite{Dinetal1999},
\cite{Coskunogluetal12} and \cite{Bil12}. In order to estimate the
parameters, we performed a test-particle integration in a Milky Way
potential which consists of a logarithmic halo, a Miyamoto-Nagai
potential to represent the galactic disc and a Hernquist potential to
model the bulge.

To calculate the galactic orbit of an open cluster, its member stars
with high membership probability should be used. Moreover, the proper
motions and radial velocities of these stars must be available. Thus,
we used only the stars whose membership probabilities are larger
than $P=\%90$ and already have radial velocities
measurements from the high-dispersion spectra by \cite{LBetal2015}. 
The number of stars in the field of NGC~6819 satisfying these 
conditions is 12. However, one of them was removed from the list 
since it is a spectroscopic binary, reducing the number to 11. The 
data collected for these stars are listed in Table~9. The columns 
of the Table~9 are organized as WOCS ID, ID in our study, equatorial 
coordinates, apparent magnitude ($V$), colour ($B-V$), proper motion 
components ($\mu_{\alpha}\cos \delta$, $\mu_{\delta}$), radial 
velocity ($V_{r}$) and the probability of membership ($P$). The 
proper motions of the stars were taken from the astrometric catalogue 
of \citet{Roesetal2010}. Mean values of the input parameters for the 
cluster's galactic orbit estimation were obtained from Table~9: 
$V_{r} = 2.51\pm 1.48$ km s$^{-1}$, $\mu_{\alpha}\cos{\delta}=-3.84\pm 4.25$ 
and $\mu_{\delta}=-2.13\pm 4.25$ mas yr$^{-1}$, and $d=2309\pm 106$ pc, 
respectively. The distance of stars was assumed to be the value found 
in this study (see Section 4.3). \cite{Wuetal2009} used almost the same 
parameters ($\mu_{\alpha}\cos{\delta}=-3.14\pm 1.01$, 
$\mu_{\delta}=-3.34\pm 1.01$ mas yr$^{-1}$ and $V_{r}=4.8\pm 0.9$ km s$^{-1}$) 
to calculate the parameters of the galactic orbit of the cluster. Galactic 
orbit of the cluster was determined within an integration time of 3 Gyr 
in steps of 2 Myr. With this integration time, the cluster completes minimum 
12 revolutions around the galactic center. Thus, the averaged orbital 
parameters can be determined reliably. We also determined the 
galactic orbits of the cluster stars in Table~9.

Representations of galactic orbits calculated for the selected cluster
stars and the cluster itself in the $X-Y$ and $X-Z$ planes are shown
in Fig.~13. In Fig.~13, $X$, $Y$ and $Z$ are heliocentric galactic
coordinates directed towards the galactic centre, galactic rotation
and the north galactic pole, respectively.

\begin{figure}[h]
\begin{center}
\includegraphics[scale=0.4, angle=0]{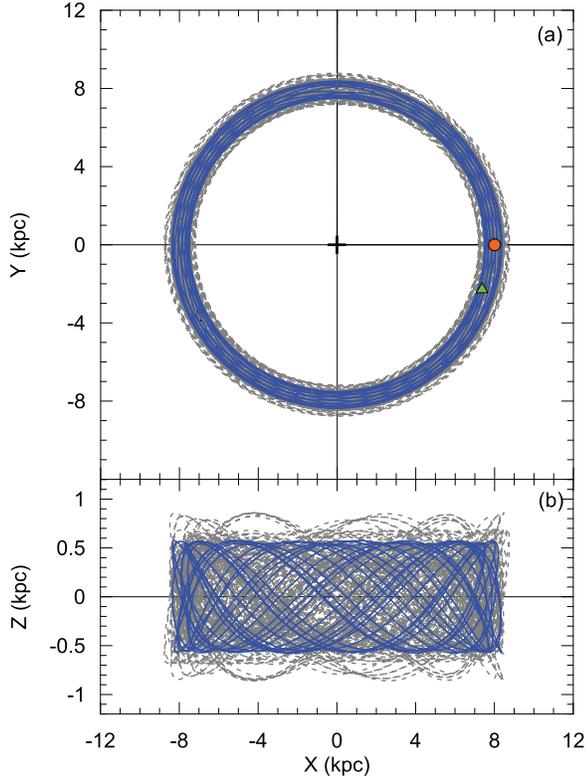}
\caption[] {\small The galactic orbital motions (grey dashed lines) of 
the 11 cluster stars with $P>90\%$, for which radial velocities are 
available, in the $X-Y$ (a) and $X-Z$ (b) planes. The cluster's mean 
orbit is indicated with a blue line. The black plus, red circle and 
green triangle symbols in panel (a) represent the galactic centre, and current 
locations of the sun and NGC 6819, respectively.}
\end{center}
\end{figure}

The cluster's apogalactic ($R_{max}$) and perigalactic ($R_{min}$)
distances were obtained as 8.42 and 7.50~kpc, respectively. The
maximum vertical distance from the galactic plane is calculated as
$Z_{max}= 580~{\rm pc}$. When determining the eccentricity projected
on to the galactic plane, the following formula was used:
$e=(R_{max}-R_{min})/(R_{max}+R_{min})$. The eccentricity of the orbit
was calculated as $e=0.06$. This value shows that the cluster is
orbiting the Galaxy with a period of $P_{orb}=142$ Myr. Although the
cluster's orbital parameters are generally in agreement with those
estimated by \cite{Wuetal2009}, the orbital period in our study is
considerably shorter than their estimate ($220.5$ Myr).
It is not clear the origin of the difference between the orbital 
periods estimated in our study and that in \cite{Wuetal2009}, while 
the other parameters are generally in agreement.

\subsection{Luminosity and mass functions of the cluster}

The relative number of stars in the unit absolute magnitude range is
termed as the luminosity function (LF). The problem in the 
estimation of the LF of an open cluster is the contamination caused 
by the field stars that are not physical members of the cluster. The 
effect of non-member stars was demonstrated in this study using the 
following procedure. First, we selected the turn-off and main-sequence 
stars with $15.5\leq V \leq 19$ mag located in a circular field of 
6 arcmin radius from the centre of the cluster and in the band-like 
region as defined in Section~3.3. There are 553 stars that satisfy 
these conditions. Then we removed the stars whose membership 
probabilities could not be estimated because of the lack of proper 
motion measurements. This selection procedure resulted in 455 stars 
with the membership probability $P>0\%$.

We calculated absolute magnitudes of the main-sequence stars selected
with this procedure using the distance modulus $\mu_{V}=12.22\pm 0.10$
mag in Table~7, resulting an absolute magnitude range $3.3\leq
M_{V}\leq 6.8$~mag. We also calculated their masses utilizing the
theoretical isochrone of the cluster. The mass range of the stars is
found as $0.765\leq M/M_\odot \leq 1.303$ for the cluster. The LFs of
NGC~6819 were estimated for the stars with the membership
probabilities of $P\geq 20\%$ ($N=381$) and $P\geq 50\%$ ($N=299$) in
order to demonstrate the effect of non-member field stars. Fig.~14
displays the LFs of the cluster.

The mass function (MF) is defined as the relative number of stars in a
unit range of mass centered on mass $M$ and represents the rate of
star creation as a function of stellar mass. For NGC 6819, we used
theoretical models provided by the PARSEC synthetic stellar library
\citep{Bresetal2012} to convert the LFs to MFs for NGC~6819. The MFs
of the cluster are shown in Fig. 15. The slope $x$ of mass function
was derived from the following linear relation: $\log
(dN/dM)=-(1+x)\log(M)+C$, where $dN$ represents the number of stars in
a mass bin $dM$ with central mass of $M$, and $C$ is a constant. We
found the slopes of the MFs to be $x=1.13 \pm 0.64$ and $x=1.70 \pm
0.78$ for the stars with the membership probabilities of $P\geq 20\%$
and $P\geq 50\%$, respectively. Since these values are roughly in
agreement, we conclude that the effect of non-member stars is not
important for this cluster. Thus, we adopt the MF slope $x=1.13 \pm
0.64$ for NGC~6819. Note that this MF slope is close to the value of
1.35 given by \cite{Salpe1955} for the stars in the solar
neighbourhood.

\begin{figure}
\begin{center}
\includegraphics[trim=0cm 8cm 0cm 0cm, clip=true, scale=0.4]{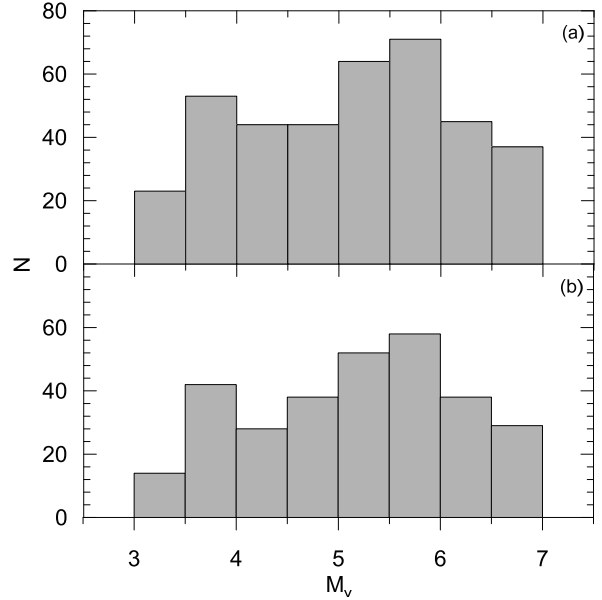}
\caption[]{\small The luminosity functions of NGC 6819 estimated
for the stars with the membership probability $P\geq 20\%$ (a) and
$P\geq 50\%$ (b).}
\end{center}
\end{figure}

\begin{figure}
\begin{center}
\includegraphics[trim=0cm 8cm 0cm 0cm, clip=true, scale=0.4]{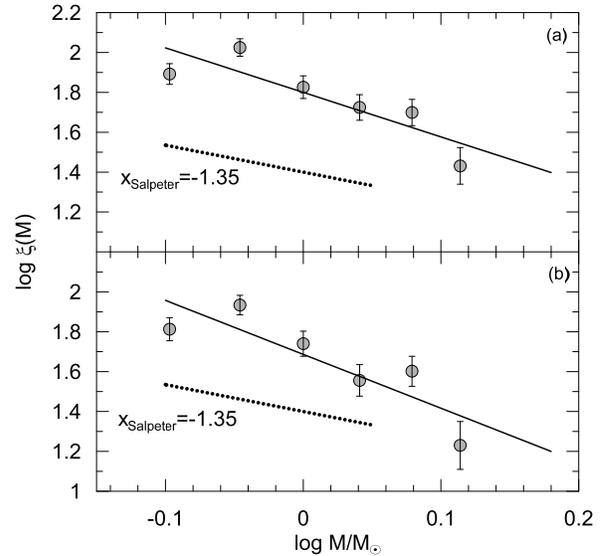}
\caption[] {\small The mass functions of NGC~6819 estimated for the
stars with the membership probability $P\geq 20\%$ (a) and
$P\geq 50\%$ (b).}
\end{center}
\end{figure}

\section{Conclusions}

In this paper, we present CCD $UBV$ photometry for the open cluster
NGC~6819 and determine structural and astrophysical parameters of the
cluster. The astrometric data of the stars were used to estimate their
membership probabilities. Additionally, we estimated galactic
orbital parameters and mass function of the cluster.

Since the astrophysical parameters of a cluster suffer from the
reddening-age degeneracy when they are simultaneously determined by
fitting the theoretical stellar evolutionary isochrones to the
observed CMDs \citep[cf.][]{Anders2004, Kingetal2005, Brid2008, deMeule2013},
independent methods developed for the determination of these
parameters are very promising to reduce the number of free
parameters. Thus, instead of using the isochrone fits, we inferred the
reddening of the open cluster NGC~6819 from its $U-B$ vs $B-V$ TCD,
while we determined the metallicity of the cluster utilizing F0-G0
spectral type main-sequence stars \citep{Cox2000} via a metallicity
calibration defined by \cite{Karetal2011}. The metallicity
 obtained from the photometry is in agreement with the one inferred
 from the spectroscopic observations \citep{LBetal2015} and is close
 to the solar value. Therefore, we assumed the metallicity of the
 cluster to be the solar value and derived the distance modula and
 age of NGC~6819 by fitting the theoretical isochrones to the
 observed CMDs keeping the metallicity and reddening fixed. 
 This method allows us to break in part the reddening-age
 degeneracy.

A comparison of Tables 1 and 6 reveals that the reddening and
metallicity obtained from the independent methods are in agreement
with those reported in previous photometric and spectroscopic studies
in general. The age and distance modulus of the cluster derived by
fitting the theoretical isochrones with the CMDs are in agreement with
previous estimates, as well. The main results can be summarized as
follows:

\begin{enumerate}

\item The central stellar density, core radius and the background
 stellar density for the cluster are determined as $f_{0}=13.18 \pm
 0.46$ stars arcmin$^{-2}$, $r_{c}=3.65 \pm 0.38$ arcmin and
 $f_{bg}=5.98\pm 0.45$ stars arcmin$^{-2}$, respectively.

\item The reddening of the cluster was determined as $E(B-V)=0.130\pm
 0.035$ mag using the $U-B$ vs $B-V$ TCD.

\item The metallicity of NGC~6819 is inferred as $[Fe/H]=+0.051\pm
 0.020~{\rm dex}$ using the UV excesses of the F and G type
 main-sequence stars of the cluster. This photometric metallicity is
 in agreement with photometric and spectroscopic metallicities
 appeared in the literature, in general.

\item The distance modula, the distance and the age of NGC~6819 were
 derived to be $\mu_{V} = 12.22 \pm 0.10$ mag, $d = 2309 \pm 106$ pc
 and $t = 2.4 \pm 0.2$ Gyr, respectively, by fitting the theoretical
 isochrones to the observed CMDs of the cluster. We demonstrated
 that the distance of the cluster is in agreement with the one
 estimated using only the RC stars of NGC~6819.

\item Precise kinematical data of the cluster allowed for an
 estimation of its galactic orbit. The cluster's apogalactic
 ($R_{max}$) and perigalactic ($R_{min}$) distances, the maximum
 vertical distance from the galactic plane, the orbital eccentricity
 projected on to the galactic plane and the orbital period were
 calculated as $R_{max}= 8.42$ kpc, $R_{min}= 7.50$ kpc, $Z_{max}=
 580~{\rm pc}$, $e=0.06$ and $P_{orb}= 142$ Myr, respectively.

\item The slope of the mass function for the cluster is derived as
 $x=1.13 \pm 0.64$. This slope value is close to the value of 1.35
 derived by \cite{Salpe1955} for the stars in the solar neighbourhood.

\end{enumerate}

\section{Acknowledgments}

Authors are grateful to the anonymous referee for his/her considerable
contributions to improve the paper. 
This work has been supported in part by the Scientific and
Technological Research Council (T\"UB\.ITAK) 113F270. Part of this
work was supported by the Research Fund of the University of Istanbul,
Project Numbers: 39170 and 39742. We thank to T\"UB\.ITAK National
Observatory for a partial support in using T100 telescope with project
number 15AT100-738. We also thank to the on-duty observers and members
of the technical staff at the T\"UB\.ITAK National Observatory for
their support before and during the observations. 
This project is financed by the SoMoPro II programme (3SGA5916).
The research leading to these results has acquired a financial grant
from the People Programme (Marie Curie action) of the Seventh
Framework Programme of EU according to the REA Grant Agreement
No. 291782. The research is further co-financed by the South-
Moravian Region. It was also supported by the grant GP14-26115P.
This work reflects only the authors views and the European Union
is not liable for any use that may be made of the information 
contained therein. This research has made use of the WEBDA, SIMBAD, 
and NASA\rq s Astrophysics Data System Bibliographic Services.

\end{document}